\documentclass{Rinton-P9x6}

\begin{document}

\title{Fiber-Optic Sources of Quantum Entanglement}

\author{Prem Kumar, Xiaoying Li, Marco Fiorentino, Paul L. Voss,\\
Jay E. Sharping, and Geraldo A. Barbosa}

\address{Center for Photonic Communication and Computing, ECE Department
\\ Northwestern University, 2145 N. Sheridan Road, Evanston, IL
60208-3118 \\ E-mail: kumarp@northwestern.edu
}


\maketitle

\abstracts{We present a fiber-based source of
polarization-entangled photon pairs that is well suited for
quantum communication applications in the 1.5\,$\mu$m band of
standard telecommunication fiber. Quantum-correlated signal and
idler photon pairs are produced when a nonlinear-fiber Sagnac
interferometer is pumped in the anomalous-dispersion region of the
fiber. Recently, we have demonstrated nonclassical properties of
such photon pairs by using Geiger-mode InGaAs/InP avalanche
photodiodes. Polarization entanglement in the photon pairs can be
created by pumping the Sagnac interferometer with two orthogonally
polarized pulses. In this case the parametrically scattered
signal-idler photons yield biphoton interference with $>90\%$
visibility in coincidence detection, while no interference is
observed in direct detection of either the signal or the idler
photons.}

\section{Introduction}
\label{intro}

Quantum entanglement refers to the nonclassical dependency of
physically separable quantum systems. It is an essential resource
that must be freely available for implementing many of the novel
functions of quantum information processing, such as database
searching, clock synchronization, teleportation, computing, and
cryptography.\cite{bennett+shor98} Therefore, the efficient
generation and transmission of quantum entanglement is of prime
importance. With the ubiquitous standard optical fiber serving as
the transmission medium and the widespread availability of
efficient active and passive fiber devices, technological synergy
between the generation and propagation components of the overall
quantum network can be achieved by deploying sources of
entanglement that rely on the nonlinearity of the fiber itself. In
order to develop such sources, we have been conducting experiments
with fiber-optic devices that can be used as building blocks for
fiber-based sources of either polarization or quadrature
entanglement. In the polarization case, our fiber-based setup for
single-photon quantum optics has the advantage of modal purity
over its crystal counterparts,\cite{kwiat99,gisin01} which would
be very important for realizing complex networks involving several
nonlinear elements.

\section{An All-Fiber Source of Quantum-Correlated Photon Pairs}
\label{afsqcpp}

In our effort to develop fiber-based sources of entanglement, we
have been investigating nondegenerate four-wave mixing in standard
dispersion-shifted fiber (DSF), wherein two pump photons scatter
through the Kerr nonlinearity of the fiber to create simultaneous
signal and idler photons. Our experiment is conducted near the
zero-dispersion wavelength of the DSF, where such scattering is
enhanced owing to phase matching of the photon wave functions. It
must be pointed out here that in a conventional
wavelength-division-multiplexed classical optical communication
line, one strives to suppress the four-wave or four-photon mixing
process, which otherwise causes cross-talk between the wavelength
channels and sets limits on the total data capacity of the
communication line.

\begin{figure}
\vspace*{-0.125in}
\centerline{\scalebox{1}{\includegraphics{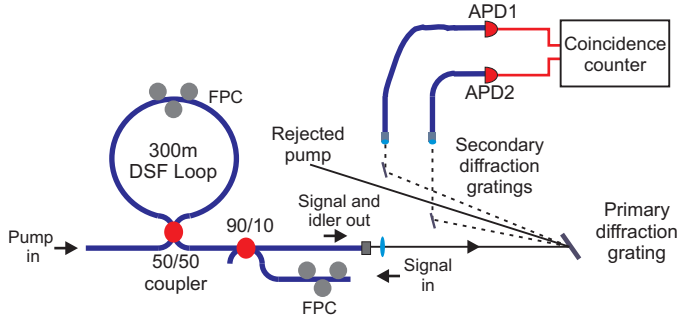}}}
\vspace*{-0.125in} \caption{Schematic of the experimental setup.
Dispersion-shifted-fiber (DSF) loop and the 50/50 coupler
constitute the NFSI. FPC, fiber polarization controller; APD,
avalanche photodiode. The signal-in port is blocked during
photon-counting measurements.} \vspace*{-0.125in}
\label{CoincSchem}
\end{figure}

As shown in Fig.~\ref{CoincSchem}, the four-photon mixing takes
place in a nonlinear-fiber Sagnac interferometer (NFSI), which we
have previously used to generate quantum-correlated twin beams in
the fiber.\cite{sharping01a} The pump is a mode-locked train of
$\simeq 3$\,ps long pulses that arrive at a 75.3\,MHz repetition
rate. The pulsed operation serves two important purposes: i) the
NFSI amplifier can be operated at low average powers and ii) the
production of the fluorescence photons is confined to well-defined
temporal windows, allowing a gated detection scheme to be used to
increase the signal-to-noise ratio. To measure the nonclassical
(i.e., quantum) correlations between the signal and the idler
photons, one must effectively suppress the pump photons from
reaching the detectors. Since a typical pump pulse contains
$\simeq 10^8$ photons and we are interested in detecting $\simeq
0.01$ signal/idler photon pairs per pulse, a pump-to-signal
(idler) rejection ratio in excess of 100\,dB is required. We
achieve this specification by sending the fluorescence photons
through a free-space double-grating spectral filter that separates
the signal and idler photons from each other and from the pump
photons not rejected by the NFSI (see Fig.~\ref{CoincSchem}).

\begin{figure}
\vspace*{-0.125in}
\centerline{\scalebox{1.1}{\includegraphics{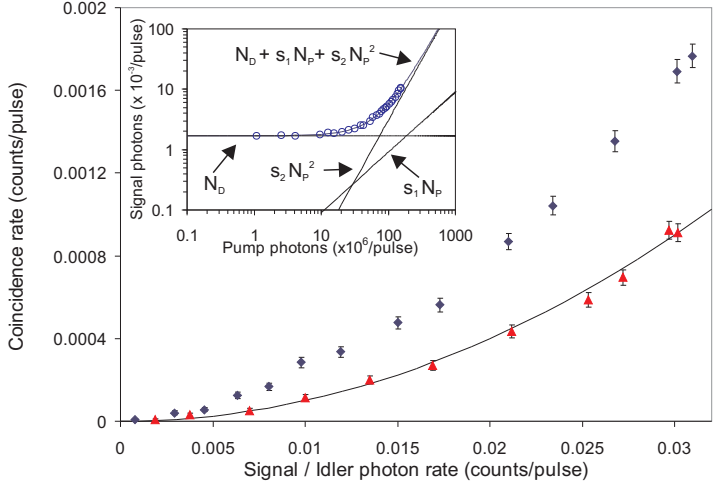}}}
\vspace*{-0.125in} \caption{Coincidence rates as a function of the
single-photon rates in two different cases: signal-idler
fluorescence produced by a pump pulse (diamonds) and signal-idler
fluorescence produced by two consecutive pump pulses (triangles).
The line represents the calculated ``accidental'' counts. The
inset shows a plot of the detected signal (or idler) photons as a
function of the injected pump photons (hollow circles). A
second-order polynomial is shown to fit the experimental data. The
contributions of the dark counts, linear scattering, and quadratic
scattering are plotted separately as well.} \label{CoincMeas}
\vspace*{-0.125in} \end{figure}

To demonstrate the quantum nature of such four-photon scattering
at the ``single" photon level, we have assembled a photon-counting
apparatus for detecting the signal and idler photons in
coincidence. This apparatus is based on commercial InGaAs
avalanche photodiodes operating in the gated Geiger mode. Use of
this apparatus has allowed us to verify the simultaneous
production of the signal and idler photon
pairs.\cite{fiorentino02b} In the inset in Fig.~\ref{CoincMeas} we
show the number of scattered photons, $N_I$ ($N_S$), detected in
the idler (signal) channel as a function of the number of pump
photons, $N_P$, injected into the NFSI. We fit the experimental
data with $N_I= N_D + s_1 N_P + s_2 N_P^2$, where $N_D$ is the
number of dark counts during the gate interval, and $s_1$ and
$s_2$ are the linear and quadratic scattering coefficients,
respectively. The fit clearly shows that the quadratic scattering
owing to four-photon mixing in the fiber can dominate over the
residual linear scattering of the pump due to imperfect filtering.
The main body of Fig.~\ref{CoincMeas} shows the coincidence
counting results. The diamonds represent the rate of coincidence
counts as a function of the geometric mean of the rates of the
signal and idler photons generated during the same pump pulse.
Dark counts have been subtracted from the plotted count rates. The
triangles in Fig.~\ref{CoincMeas} represent the measured
coincidence rate as a function of the signal-photon count rate
when the signal is delayed with respect to the idler by one pulse
period. For two independent photon sources, each with a count rate
$R_S \ll 1$, the ``accidental'' coincidence rate $R_C$ is given by
$R_C=R_S^2$, regardless of the photon statistics of the sources.
This quadratic relation is plotted as the solid line in
Fig.~\ref{CoincMeas}, which fits the delayed-coincidence data
(triangles) very well. These measurements then show that while the
fluorescence photons produced by the adjacent pump pulses are
independent, those coming from the same pump pulse show a strong
correlation, which is a signature of their nonclassical behavior.

With 75\,MHz pump-pulse rate, we are now routinely detecting over
1500 photon pairs/s, which we expect to push up by a factor of 20
with further refinements in the detection apparatus. Use of higher
repetition rate pump pulses (over 10\,GHz rate is possible with
mode-locked fiber lasers) has the potential of increasing this
rate by another two orders of magnitude. As we demonstrate in
Sec.~\ref{fospe} below, polarization-entangled photon pairs can
then be obtained by polarization multiplexing the signal and idler
photons from two orthogonally-polarized pumps.

\section{A Fiber-Optic Source of Polarization Entanglement}
\label{fospe}

In this section we present the first, to the best of our
knowledge, all-fiber source of polarization-entangled photon
pairs. This source has the advantage of modal purity over its
crystal counterparts,\cite{kwiat99,gisin01} which is very
important for realizing complex networks involving many quantum
operations.

\begin{figure}
\vspace*{-0.25in}
\centerline{\scalebox{0.58}{\includegraphics{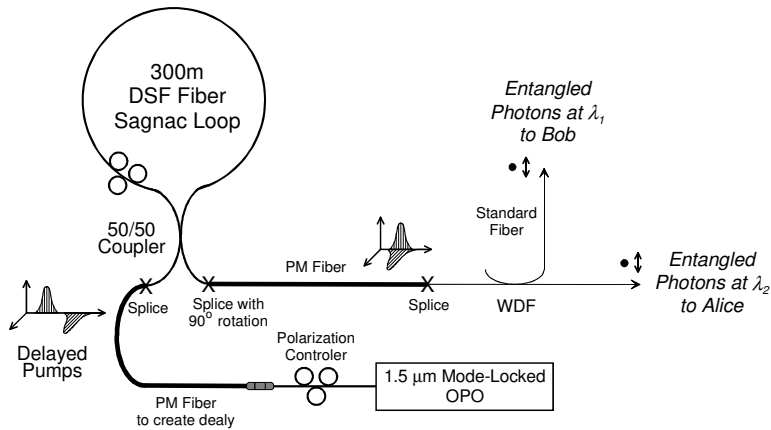}}}
\vspace*{-0.125in}\caption{Schematic of an all-fiber experiment to
create polarization-entangled photon pairs. In our current
implementation, the relatively-delayed, orthogonally-polarized
pumps at the input are created with bulk-optic components. The
delay at the output, however, is removed by use of a
polarization-maintaining (PM) fiber, as shown. Similarly, in our
current setup the polarization-entangled photon pairs generated in
the Sagnac loop are separated at the output with a bulk-optic
wavelength-dependent filter (WDF). An approximately 400-m long
all-fiber polarization interferometer is formed between the
entrance surface of the input PM fiber and the exit surface (just
before the WDF) of the output PM fiber.} \vspace*{-0.125in}
\label{PolEntSchem}
\end{figure}

\begin{figure}[t]
\vspace*{-0.125in}
\scalebox{.65}{\includegraphics{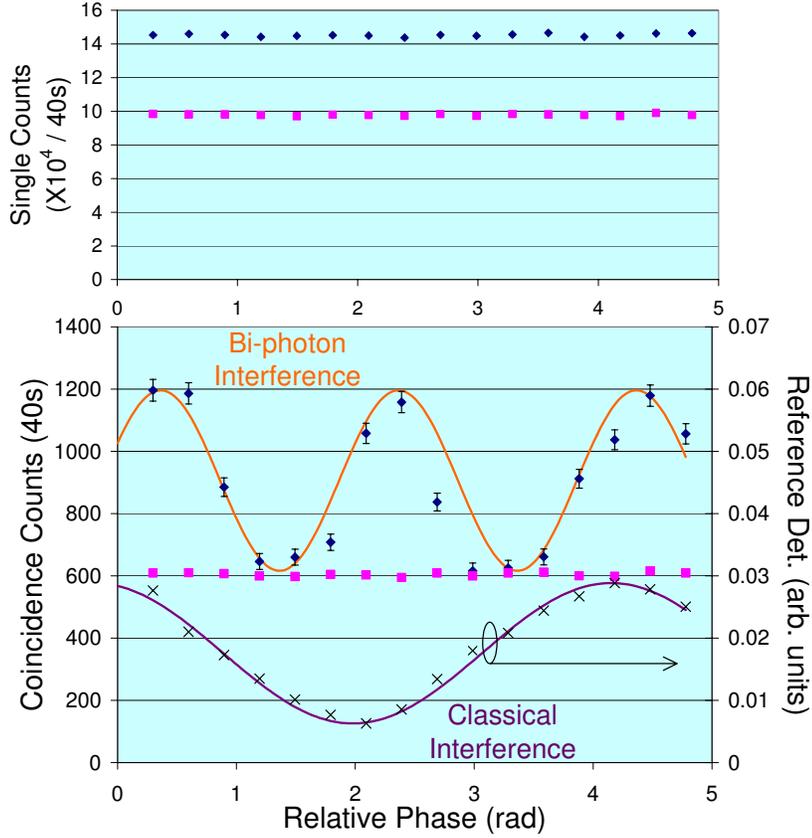}}
\vspace*{-0.125in} \caption{Top Plot: Single-photon counts
registered by the signal (diamonds) and the idler (squares)
detectors as the phase between the two pump pulses is varied. No
phase dependence is observed in either detector's count rate.
Bottom Plot: Coincidence-photon counts (diamonds) registered by
the signal and idler detectors as the phase between the two pump
pulses is varied. The squares represent accidental coincidences
that are measured due to less-than-unity detection efficiency and
dark counts in the detectors. After subtraction of the accidental
counts, the resulting fringe visibility is $> 90\%$. The period of
the fringes is half that of the classical interference fringes
(crosses, right scale), which were simultaneously monitored by a
reference detector during the coincidence measurements. The solid
lines are best fits to the expected sinusoidal dependencies.}
\vspace*{-0.125in} \label{BiPhotonInt}
\end{figure}

As demonstrated in Sec.~\ref{afsqcpp} above, the parametric
fluorescence accompanying nondegenerate four-wave mixing in
standard optical fibers is an excellent source of
quantum-correlated photon pairs.\cite{fiorentino02b} The quantum
correlations arise because the parametric fluorescence is a result
of two pump photons scattering through the Kerr nonlinearity to
create simultaneous signal and idler photons. In order to create
polarization-entangled photon pairs, we take advantage of the fact
that the parametric fluorescence in standard fiber is
predominantly co-polarized with the pump. When the Sagnac loop is
pumped with two delayed orthogonally-polarized pulses, see
Fig.~\ref{PolEntSchem}, the down-converted signal and idler
photons originating from the two pulses are co-polarized with the
corresponding pumps. However, when the delay distinguishing the
down-converted photons is removed and the signal/idler photons are
wavelength separated, then a polarized single-photon detection
event at the output, in either the signal or the idler arm, cannot
determine which pump-pulse produced the photon. This
indistinguishability results in polarization entanglement of the
signal and idler photons, which we demonstrate by means of
interference in coincidence-photon counting of the polarized
signal and idler outputs. The coincidence-rate should vary
sinusoidally as the relative phase between the two pump pulses is
scanned.

Similar nonclassical interference has previously been demonstrated
by Ou~{\em et al.} with use of two separate $\chi^{(2)}$-crystal
based spontaneous parametric down-converters.\cite{owm89,owzm90}
The observed visibility was 40\%. The two-interferometer setup in
their experiment was of the Mach-Zehnder type, whereas in our
experiment the two-interferometer is formed by temporally
multiplexing an all-fiber polarization interferometer. This
interference effect is based on pairs of photons (biphotons)
rather than single photons, which led Ou {\em et al.} to the
remark: ``Indeed it is possible to modify Dirac's famous statement
and argue that pairs of photons are interfering with themselves in
these experiments."\cite{owzm90}

A summary of the details of our observations is given in the
caption of Fig.~\ref{BiPhotonInt}. We observe a visibility of
$>90\%$ in the interference fringes that occur in coincidence
counting. Classically, no interference is expected in such
coincidence counts.

\section*{Acknowledgments}
This work was supported by the Department of Defense
Multidisciplinary University Research Initiative (MURI) program
administered by the Army Research Office under Grants
DAAD19-00-1-0177 and DAAD19-00-1-0469, and by the U. S. Office of
Naval Research under Grant N00014-91-J-1268.

\end{document}